\documentclass[12pt,a4paper]{article}

\usepackage{amsmath}
\usepackage{amssymb}

\begin{document}

\newcommand{\mb}{\ensuremath{\mathbf}}
\newcommand{\mc}{\ensuremath{\mathcal}}
\newcommand{\vc}{\ensuremath{\vec}}
\newcommand{\dv}{\ensuremath{\partial}}
\newcommand{\tb}{\textbf}
\newcommand{\txt}{\textit}
\newcommand{\txl}{\textsl}
\newtheorem{ipo}{Ipotesi}
\newtheorem{corol}{Corollario}

\newtheorem{nb}{Nota bene}
\setcounter{section}{0}

\renewcommand{\thesection}{\large\arabic{section}.}
\renewcommand{\thesubsection}{\thesection \arabic{subsection}}
\newtheorem{propr}{Propriet\`{a}}
\newtheorem{prop}{Proposition}
\newtheorem{oss}{Osservazione}
\newtheorem{lem}{Lemma}
\newtheorem{cri}{Criterio}
\newtheorem{cond}{Condizione}

\newcounter{n1}
\newcounter{n2}
\newcounter{n3}
\newcounter{n4}
\newcounter{n5}
\newcounter{n6}
\newcounter{n7}
\newcounter{n8}
\newcounter{n9}

\newenvironment{ls1}{\begin{list}{\small(\roman{n1})\normalsize}
{\usecounter{n1}}}{\end{list}}
\newenvironment{ls3}{\begin{list}{\tb{(\txl{40.\alph{n3}})}}
{\usecounter{n3}}}{\end{list}}
\newenvironment{ls4}{\begin{list}{(\txl{L.\alph{n4}})}{\usecounter{n4}}}{\end{list}}
\newenvironment{ls5}{\begin{list}{(\txl{R.\alph{n5}})}{\usecounter{n5}}}{\end{list}}
\newenvironment{ls6}{\begin{list}{\small\txl{\alph{n6}})}
{\usecounter{n6}}}{\end{list}}
\newenvironment{ls7}{\begin{list}{\small\arabic{n7}.\normalsize}{\usecounter{n7}}}
{\end{list}}
\newenvironment{ls8}{\begin{list}{\Alph{n8}.\normalsize}{\usecounter{n8}}}{\end{list}}
\newenvironment{ls9}{\begin{list}{\small\Roman{n9}\normalsize}{\usecounter{n9}}}
{\end{list}}


\def\reali{\hbox{\rm I\hskip-2pt\mb R}}
\def\realipunto{\hbox{\rm I}\hskip-2pt \dot {\mb R}}
\def\complessi{\hbox{\rm I\hskip-5.9pt\mb C}}
\def\razionali{\hbox{\rm I\hskip-5.7pt\mb Q}}
\def\interi{\hbox{\mathrm{Z\hskip-8.2pt Z}}}
\def\naturali{\hbox{\mathrm I\hskip-5.5pt\bf N}}

\def\dsec#1#2{{{\partial^2 #1 } \over {\partial #2^2}}}
\def\dpar#1#2{{{\partial #1 } \over {\partial #2}}}
\def\modulo#1{\left| #1 \right|}
\def\norma#1{{\left\| #1 \right\|}}
\def\eps{\varepsilon}


\def\bsk{\vspace {1.5 cm}}
\def\msk{\vspace {0.5 cm}}
\def\ssk{\vspace {0.2cm}}
\def\ni{\noindent}
\def\ac{\par\noindent}


\font\titfnt=cmbx10 at 14.40 truept
\font\tentex=cmtex10
\font\ninerm=cmr9
\font\eightrm=cmr8
\font\sixrm=cmr6

\def\R{{\mc R}}
\def\xn{x_1, \ldots ,  x_n}
\def\yn{y_1, \ldots ,  y_n}
\def\ddxi{{\partial \over \partial x_i}}

\def\<{\langle}
\def\>{\rangle}

\hspace {8 cm}   IFUP-TH 2008/08

\bsk
\centerline{A. Cintio}
\ssk  
\centerline{Dipartimento di Fisica and INFN, Pisa, Italy}
\msk
\centerline{G. Morchio}
\ssk
\centerline{Dipartimento di Fisica and INFN, Pisa, Italy}
\bsk
\centerline{\bf Sum rules and density waves spectrum}
\ssk
\centerline{\bf for non relativistic fermions}

\bsk\ni
\abstract{Frequency sum rules are derived 
in  extended quantum systems of non relativistic fermions 
from a minimal set of assumptions on dynamics in infinite 
volume, for ground and thermal states invariant
under space translations or a lattice subgroup. 

\ni
For the jellium  Coulomb model, they imply the one 
point result for the plasmon energy spectrum in 
the zero momentum limit. 

\ni
In general, the density waves energy spectrum is shown to converge, in 
the limit of large wavelenght, to a point measure at 
zero frequency, for any number of fermion fields and potentials 
$V^{(ij)}$ with integrable second derivatives.

\ni
For low momentum, $ < \omega^2 (k) >  \sim k^2 $ 
for potentials with $ r^2 \, \dv_i \dv_j V $ integrable, 
$ < \omega^2 (k) > \sim  k^{\alpha - d + 2}$  for  potentials 
decaying  at infinity  as  $1/r^\alpha$, $ d-2 < \alpha < d $, 
$d$ the space dimensions. 

\ni For one component models with short range interactions, 
the fourth momentum of the frequency is expressed, 
at lowest order in $k$, purely in terms of the three point 
correlation function of the density.} 

\bsk\ni
Math. Subj. Class.: 82B10, 82B21

\ssk\ni 

\newpage                                 

\section {Introduction}

The physics of charge density waves in extended systems was
understood in ref.\cite{Tonks-langmuir1929} in terms of classical 
charge configurations giving rise to slowly varying electric 
fields, resulting in \txl{plasma oscillations} 
with frequency $\omega_p^2 = e^2 \rho / m $ 
for infinitely extended systems of particles of mass $m$, 
charge $e$ and density $\rho$.

In solid state physics such an analysis applies \cite{Kubo:book}
on the basis of suitable simplifications, 
in particular of a \txl{random phase approximation} 
\cite{Pines:book} in the analysis of charge density 
correlation functions.  Plasma oscillations correspond to a
single point frequency spectrum, $\omega = \omega_p$, for the charge 
density correlation function at zero momentum, and it is 
not clear whether neglected terms may spoil such pure point result, even 
in the case of only one kind of charged particles in a neutralizing 
uniform background (jellium model).

The control of the energy spectrum associated to density waves also plays 
a crucial r\^ole in the the theory of quantum liquids, in particular in the 
discussion of superfluidity, 
\cite{Feynman1954} \cite{Feynman1955} 
\cite{Pines-Nozieres:book}; very similar problems appear in the discussion of 
the current commutators at low momentum which are at the basis of the Kubo and 
Landau approaches to superconductivity \cite{Landau1941} \cite{Strocchi:book}.

In ref. \cite{Morchio-strocchi1986} the relevant spectrum in the jellium 
Coulomb model was derived, and shown to consist only of the plasma frequency, 
through an analysis of Galilei transformations, which are 
spontanously broken in the jellium model, as in any nonrelativistic 
system at non--zero density.

The derivation employed a \txl{generalized Goldstone theorem}, which gives 
the energy spectrum associated to spontaneosly broken symmetries in terms of
commutators between generators of symmetries and order parameters. 
Such commutators are time independent in the ordinary
Goldstone case, but not in the presence of sufficiently long range
interactions, the Coulomb potential in the jellium case.
Similar results have also been obtained in refs.
\cite{Broidioi-verbeure1993} through an
analysis of operators describing long wave charge fluctuations in the 
jellium and other models.

The Galilei analysis is complicated by the lack of invariance under
space translations of the density of the Galilei generator and by 
short distance problems associated to the singularity 
of the Coulomb potential at the origin. The analysis can in fact be 
simplified by focusing on charge density commutators, 
a tecnique discussed e.g. in refs. \cite{Puff1965}, \cite{Stringari1992},
\cite{Stringari1994}. 
On one side, such commutators provide an alternative derivation of 
the spectrum associated to Galilei transformations, on the other they 
have convenient positivity properties and simple evolution equations.
Moreover, as we shall see, the charge commutators analysis applies 
to a large class of systems and allows for the derivation of exact 
relations for (density waves) energy spectra, also at non--zero momentum.

The aim of this paper is to present a self contained derivation 
of the plasma spectrum at zero momentum in the jellum
model, based on the analysis of charge commutators, to extend the analysis 
to the case of spontaneous breaking of translation to a lattice subgroup 
(which is relevant in presence of Wigner crystallization \cite{Ceperley}, 
\cite{Ciccariello}) and to derive, by the same methods, general relations for 
the frequency spectrum at low momenta of density waves in non
relativistic fermion systems with short and long range potentials.

The use of sum rules to obtain information on the density waves frequency
spectrum seems to have been prevented by results on the divergence of momenta
of order $ \geq 3$ \cite{Family1975}. However, the divergence of the third momentum
only appears for singular (delta or hard core) potentials and the divergence
of the fifth momentum in the Coulomb case depends on the singularity
at the origin of the potential. More generally, the (perturbative) results of
\cite{Family1975} indicate that no momentum diverges for regular ($C^\infty$) 
potentials, a result which also follows from our analysis. 
In the following, the Coulomb potential will be regularized at the origin and 
the limit $ k \to 0$ of the energy spectral measure will be shown to be 
independent of the regularization.
  
We remark that the result on the plasma spectrum at zero momentum 
does not follow from the sum rules for the frequency momenta 
up to the third, nor by the use of the \lq\lq perfect screening sum rule\rq\rq, 
concerning $\<\omega^{-1}\>$ \cite{Pines-Nozieres:book}, unless  
a one point approximation is assumed for the frequency spectrum; 
however, the reduction of the specturm to a single point is the key 
result, which involves the fifth momentum and holds, in our analysis, 
only in the limit $k \to 0$.

Besides the analysis of the plasma spectrum and of the fifth frequency 
momentum, we reobtain the third momentum sum rules of refs.\cite{Puff1965} 
\cite{Forster1969} \cite{Stringari1992}, clarifying and symplifying
their derivation, extending them to states with discrete 
translation symmetry, and deriving exact consequences on energy spectra, 
in particular in the case of long range potentials with a faster decay 
with respect to the Coulomb potential. 
Our results do not require the explicit construction of ground and
thermal states and may also shed light on the problems and alternatives 
which appear in their analysis \cite{Lieb2005}, \cite{Disertori2007},
\cite{BelitzKirk1}, \cite{BelitzKirk2}.

We consider infinite systems of nonrelativistic fermions 
described by the canonical anticommutation relations (CAR) 
algebra, in $d$ space dimensions, in particular $d=2,3$. 
The time evolution is assumed to be given, through equal time 
commutators, by a free Hamiltonian and 
interaction potentials $ V(|x|) $.
We consider states $\Omega_\beta$, with nonzero mean particle 
density $\<\rho\>$, invariant under space translations or under a 
lattice subgroup of them, time evolution and parity, satisfying 
either the spectral condition, i.e. positivity of the energy in the 
resulting representation of the CAR algebra, or the KMS condition 
at inverse temperature $\beta$.

We denote by ${\mc F}(f)$ or by $\tilde f$ the Fourier transform of 
$f$, by $f*g$ the convolution of $f$ and $g$, by $A(f)$ the smearing of the
distribution $A(x)$ with $f$; unless specified differently, 
test functions belong to the Schwartz space ${\mc S} $ 
of infinitely differentiable functions of fast decrease.
For simplicity, the Planck constant (divided by $2 \pi$) is omitted 
and the variable $\omega$ is used for the spectrum of the Hamiltonian;
sum over repeated indexes is implicit and $dx$ will denote $d^d x$. 
Our results are the following:

\ssk\noindent
i) {\bf one--point plasma spectrum for jellium at zero momentum:} 

\noindent
In the jellium Coulomb model, i.e. for non--relativistic 
fermions in three space dimensions with interaction  
$V(x) = e^2/ 4 \pi |x| \, * \, \gamma(|x|)$, 
$\gamma \in {\mc S}$ real, $\int \gamma (x) \, dx = 1$, 
we consider the  expectations  
$$ d\mu_f (\omega) =  \int \< \rho(\bar f)  
 dE(\omega) \, \rho(f)\> \ \ ,  $$
with $f(x)= \exp{ikx} \, \alpha_R(x)$ (see eq.(\ref{fr}),  $dE(\omega) $ the   
spectral measure of the Hamiltonian, 
$\< \ \>$ the expectation on a thermal 
or ground state, invariant under  
space translations or a lattice subgroup of them and parity.
The normalized positive measures
$$
d \nu_f (\omega) \equiv N_f \, \omega \, 
(d \mu_f (\omega) - d \mu_f (-\omega))  \ \ .
$$
converge, for $\alpha_R \to 1$ and then $k \to 0$, to 
\begin{equation}  
1/2 \; (\delta(\omega - \omega_p) +
\delta (\omega + \omega_p) ) \ \  , \ \ \omega_p^2 = e^2 \< \rho \> / m 
\label{Plasmon}
\end{equation}
independently of the regularization $\gamma$. 
The derivation relies purely on Newton equations, 
Gauss' law at large distances and invariance under 
(a lattice subgroup of) translations.

\ssk\noindent
ii) {\bf Goldstone spectrum for potentials with integrable second derivatives:} 

\noindent
For any number of fermion fields and all regular potentials $V^{(kl)}$
with $\dv_i \dv_j V^{(kl)}(x)$ integrable,
the above measures $d \nu_f$, defined by the total density
$\rho = \sum_i \rho_i$,  converge to $\delta(\omega)$, 
correponding to a \txl{Goldstone spectrum} 
(also associated to the spontaneous breaking of the Galilei charges
$ \sum_i \int dx \, \rho_i(x) x_k $ ).

\ssk\noindent
iii) {\bf dependence of the spectrum on the decay properties of the potential:} 

\noindent
In the same cases as in ii) and with the same notation, 
for $R \to \infty$, i.e $f \to \exp{ikx}$,
$$ \int \omega^2 \, d\nu_f (\omega)   \to   c \, |k|^2 \, (1+o(|k|)) \ \  $$
if $ x^2 \dv_i \dv_j V(x)$ is integrable; 
$$0
\int \omega^2 \, d\nu_f (\omega)   \leq   c \, |k|^\beta \, 
$$
if $ |x|^\beta \dv_i \dv_j V(x)$, $0 < \beta <2$, is integrable,
with an equality, up to $o(|k)$, for one fermion field and 
$V(x) \sim 1/ |x|^\alpha$, $ \alpha = \beta + d -2 $.
In particular, $< \omega^2 (k) > = O(|k|)$ for jellium Coulomb 
systems (i.e. $V \sim 1/|x|$) in two space dimensions.

\ssk\noindent
iv) {\bf fourth momentum of the spectrum at the order $k^2$:} 

\noindent
For one fermion field, regular potentials of fast 
decrease and states invariant under space
translation, rotations and parity the fourth momentum of 
$d \nu_f (\omega)$ converges, for $f \to \exp{ikx}$, to 
\begin{equation}  
k^2/\< \rho \> m^2   
\; \int dx \, dy \; W(x,y) \, \< \rho(0) \rho(x) \rho(y) \> \; + \;
O(k^4) \ \  ,  
\label{fourth}
\end{equation}   
with $W(x,y)$ computed in Sect.6 in terms of the potential.
In particular, $< \omega^4 (k) >$ is not of the same order as 
$< \omega^2 (k) >^2$, as it would follow from a single quasi-particle 
interpretation.

\ssk
In Sect. 2 the mathematical framework is specified, in terms of  
correlation functions for infinite systems, with time derivatives 
given by appropriate commutators with local hamiltonians.
Energy spectra are expressed in terms of time derivatives of 
commutators; convergence to delta functions of the spectral 
measures in the limit of zero momentum follows from 
relations between frequency momenta.

In Sect. 3 equations of motion, commutators and low 
momentum expansions are discussed in general, together with 
their implications on energy--momentum spectra for short 
range interactions.

In Sect. 4 the expression for  $ <\omega^2 (k)>$ is  
shown to result, for long range potentials, in the above 
dispersion relations. 

In Sect. 5 the fourth momentum of the frequency 
is computed in the jellium model in the
limit $k \to 0$, implying the plasma frequency result. 

In Sect. 6 the first term in the $k$ expansion of 
$<\omega^4 (k)>$ is expressed in terms of the three point function
of the density for regular short range potentials.

\section{Commutators and energy spectra}

We consider infinite systems, described by fermion fields 
$\psi_i^{*}(f), \psi_i(f)$, 
generating an ACR algebra \cite{Bratteli-robinson2}, with
\begin{equation} 
[ \psi_i^{*} (f) , \psi_j (g)]_{+} = \delta_{ij} 
\int f(x) g(x) \, dx  \ \ \ \ , \ \ \
 [ \psi_i (f) , \psi_j (g)]_{+} = 0 \  , \label{ACR}
\end{equation}
$dx$ staying, here and in the following, for $d^dx$ in $d$  space
dimensions. The addition of spin indexes leads to 
minor changes, which will be indicated when relevant.  
The results of Sects. 3 and 4 hold for any number of fermion fields, 
those of Sects. 5 and 6 in the case of only one fermion field.

We consider representations defined (through the GNS construction
\cite{Bratteli-robinson2}, \cite{Sewell:book}) by states invariant under 
(a lattice subgroup of) space translations, with a finite number of particles 
when restricted to finite regions, i.e. defining locally Fock
representations. 
This allows for the use of variables in the weak closure of the ACR algebra 
in the Fock space, e.g. bounded functions of the operators $\rho(f)$, 
(density operators integrated with regular functions), 
and in fact we will work, in the spirit of Wightman 
theory \cite{Streater-wightman}, with unbounded 
field operators like $\rho(f)$. 

In the presence of interactions, the construction of the dynamics of such 
systems in terms of automorphisms of the ACR algebra is not completely 
under control; in the case of lattice  spin systems \cite{Robinson1968},  
integrability of the interaction implies norm convergence of
finite volume dynamics and stability of the quasi--local algebra, while 
the results of \cite{Morchio-strocchi1985} and \cite{Morchio-strocchi1986}
imply the necessity of weaker convergence and larger algebras in the case 
of Coulomb interactions.

For our purposes, it is enough that dynamics exists as a group of 
automorphisms of an algebra containing the ACR algebra and that expectation
values of time derivatives are given, at zero time and for suitable
variables, by limits of commutators with hamiltonians associated to 
a sequence of bounded regions invading the space.
We end therefore with the following assumptions, in the spirit of 
Wightman theory:

\ssk\noindent
A) The correlation functions at all times of the fermion fields and of 
their Wick ordered polynomials are distributions in the space 
variables, continuous in time as distributions, invariant under 
time translations and space translations, or under a lattice subgroup 
of the latter, satisfying Wightman positivity and
therefore defining operator valued distributions on the 
invariant Wightman domain, with fundamental vector $\Psi_0$ . 
The corresponding state (invariant under time and space, or lattice, 
translations) will be denoted by $\Omega$, expectations on 
$\Omega$ by $ \< \  \> $
In the space translation invariant case, the unitary groups 
implementing space and time translations are then strongly continuous; 
their generators will be denoted by $P$ and $H$, and the corresponding 
joint spectral measure by $dE(k, \omega)$. For states invariant under a
lattice subgroup of space translations, $dE(k, \omega)$ will denote
the spectral measure associated to time and lattice translations, with
$k$ in the fundamental cell of the reciprocal lattice; $dE(\omega)$ will
denote the spectral measure of $H$.

\ssk\noindent
B) $\Omega$ will be assumed to be a ground or a thermal state, i.e. 
to satisfy either positivity of the energy $H \geq 0$, or the KMS 
conditions 
\begin{equation} 
\Omega_\beta \, (A \, B(t)) = \Omega_\beta \, ( B(t - i \beta) \, A) 
\end{equation}
with $ B(t - i \beta)$ defined by analytic continuation of the 
correlation functions.
The correlation functions at equal time will be assumed to be bounded, 
after smearing, with respect to translations of space variables, to 
satisfy the cluster property and to be invariant under parity.  
Invariance under rotations will be assumed only to simplify the results, 
in the space translation invariant case.

\ssk\noindent
C) The correlation functions of time derivatives, at equal times, of Wick 
polynomials $\Phi(x)$  are given by the infinite volume limit of 
the correlation functions of their (multiple) commutators with the local 
hamiltonians
$$
(-i)^n \, d^n/dt^n \Phi = [H \ldots [H , \Phi] \ldots ] =
$$
\begin{equation} 
=  \lim_{L_i \to \infty} \lim_{R_j \to \infty} 
[ H_{L_1 , R_1} [ \ldots [H_{L_n , R_n} , \Phi] \ldots ] ] \label{ddtn}
\end{equation}
with
$$ H_{L,R} \equiv H^0_R + H^\rho_R + H^{int}_{L,R} \equiv $$
$$
\equiv 1/2m \int dx \ \alpha_{R}(x)\, \dv_{i} 
\psi^{\ast}(x)\,\dv_{i}\psi(x)\ + 
\mu  \int dx \ \alpha_{R}(x)\,  \psi^{\ast}(x)\, \psi(x)\ + 
$$
\begin{equation} 
+  1/2 \int dx \, dy\ \alpha_{R}(x) \, \alpha_{R}(y) \,
\psi^{\ast}(y)\,\psi^{\ast}(x)\,V_{L}(x-y) \, \psi(x)\, \psi(y) \ \ ,
\label{HLR}
\end{equation}
with $V_{L}(x)=\alpha_{L}(x)\; \gamma * V(|x|)$,  
$\gamma(|x|) \in {\mc S}$ 
real, $\int \gamma  \, d^dx = 1$, 
and 
\begin{equation} 
 \alpha_{R}(x) \equiv \alpha(|x|/R)  \ , \ \ \alpha \, \in {\mc S} \ , \ \
\alpha(x) \geq 0 , \ \  \alpha(0) = 1 \  ,  \ \ 
\alpha(x) = 0 \ \  \forall x \geq 1  
\label{fr}
\end{equation}
and similarly for more than one fermion field; 
the limits  $R_j \to \infty$ exist by locality of the ACR relations;
existence, and independence of the order, of the limits 
$L_i \to \infty$ follows from integrability of $V(x)$ and 
boundedness of the correlation functions in the above sense.

For gauge invariant variables, the only ones to be considered in the 
following, the second term in eq. (\ref{HLR}), as well as a possible term 
$ \rho \, V_L * \<\rho\> $, corresponding to an interaction with a 
uniform background, are irrelevant in the above commutators; 
moreover, the integrability condition applies to $\dv_i \, V(x)$, since  
only derivatives of $V$ appear, as a consequence of the vanishing 
of the commutators between gauge invariant
variables and the integral of the charge density. 

For Coulomb interactions, in order to perform the limits 
$L_i \to \infty$, the truncated correlation functions will be 
assumed to decay, after smearing with test functions, as 
$ \prod_{i=1}^{n-1} |x_{\sigma(i)} - x_{\sigma(i+1)}|^{-1 -\eps}$ 
for all permutations $\sigma$, for some $\eps > 0$.
In general, similar decay properties are required in the absence of
integrability of $\dv_i \, V$.
The ultraviolet regularization is necessary, in general, 
for the existence of frequency momenta. It will be omitted in the notation, 
writing $V$ for $\gamma * V$.

\ssk
In the above framework, we will derive constraints on energy spectra which
arise directly from the equations of motion. Goldstone theorem can be regarded 
as one of them, and follows in fact \cite{Morchio-strocchi1985} 
from invariance of the equation of motion under a symmetry which 
commutes with space translations and is spontaneously broken, 
under sufficient locality properties of the time evolution. 
For long range interactions, the latter property may fail, and in this case 
the proof of Goldstone theorem gives a relation between an energy 
spectrum at zero momentum and the time dependence of the expectation 
value of the commutator between an order parameter $A$ and a charge 
operator $Q_R$, in the limit $R \to \infty$ 
\cite{Swieca1967} \cite{Morchio-strocchi1985}

In this paper, similar information on energy spectra will be obtained 
from an analysis of  time derivatives of commutators of the form 
$ \< [A^*(x) , A(y,t) ] \> $.
If $ A \; \Psi_0$ is in the domain of $H^n$, then
$$ (-i)^n d^n/ dt^n \, \<A^*   A(t)\>_{t=0} \  = 
 \int  \omega^n  \, \<A^* \, dE(\omega) A \>  \ \ . $$
The energy spectral measure of the state $ A \Psi_0$ will be 
denoted by
$d\mu_A (\omega) \equiv  \<A^* \, dE(\omega) A\>  $;
the measures
\begin{equation} 
d \nu_A (\omega) \equiv \omega \, (d \mu_A (\omega) - d \mu_{A^*} (-\omega))
\label{dnu}
\end{equation}
satisfy, under the above domain conditions, 
\begin{equation} 
 \int d\nu_A (\omega)  \; \omega^n = 
   (-i)^{n+1} d^{n+1}/ dt^{n+1} \, \<[A^*,   A(t) ]\>_{t=0} \ \ , 
\label{momenti}
\end{equation}
For ground and thermal states the measures $d\nu_A$ are positive. If 
$ d \mu_{A^*} (\omega)  = d \mu_A (\omega)$, which will follow in our case
from parity invariance, they are even and determine $d\mu_A$ up to 
$\delta(\omega)$:

{\lem
If $\Omega$ is a ground or KMS state and $A$ hermitean, 
$d \nu_A (\omega)$, defined by eq.(\ref{dnu}), is positive.
If $ d \mu_{A^*} = d \mu_A $, then $d\nu_A$ is even and
determines $ d\mu_A$ apart from multiples of $\delta(\omega)$; 
in this case, 
if $d\nu_{A_n}$ converge as measures, $d\mu_{A_n}$ converge 
as distributions, apart from $\delta(\omega)$ terms}.

\ssk\ni
Proof: For ground states, $d\mu (\omega)$ has positive support, so that 
$ d\nu$ is clearly positive and determined up to $ c \, \delta$.
For KMS states, $d\mu(-\omega) = \exp{(- \beta \omega)} \; d\mu(\omega) $, so that
$$ d\nu(\omega) = \omega \, (1 - e^{- \beta \omega}) \; d\mu(\omega) \ \ , $$ 
which implies positivity of $d\nu$ and uniqueness of $d\mu$ apart from
multiples of $ \delta (\omega)$.
Convergence in the sense of distributions of the odd part of $d\mu_{A_n}$ 
follows from eq.(\ref{dnu}) and implies, by the ground state or the KMS condition, 
convergence of $d\mu_{A_n}$ as distributions, on test functions which vanish
at $\omega = 0$. $\Box$

\ssk
Energy spectra as functions of space momentum (pseu\-do--mo\-men\-tum in 
the case of lattice invariance) depend on
commutators of the form 
$ \<[A^*(x,0) A(y,t)]\>$; for $A$ hermitean and of definite 
parity, and $\bar f(x) = f(-x)$, i.e. $\tilde f$ real, 
parity ($P$) invariance of the dynamics and of $\Omega$ implies
\begin{equation} 
  \< A(f,0) \, A(f,t)^* \> =  \< P \, A(f,0)^* \, A(f,t) \, P \> = 
  \<  A(f,0)^* \, A(f,t)  \> \ \ , 
\end{equation}
i.e. 
$ d \mu_{A(f)^*} (\omega)  = d \mu_{A(f)} (\omega)$,
so that Lemma 1 applies and the Fourier transform $$d\nu_{A(f)} (\omega)
\equiv {\mc F} \, ( -i d/dt \, \< [ A(\bar f,0), A(f,t) ] \> \, ) $$  
is a positive even measure. 
For states invariant under space translations, 
\begin{equation} 
\<[A(f,0)^*, A(f,t)]\> \,  =  
\int \< A \, (dE(k,\omega) - dE(k,-\omega)) \, A\> \, \tilde f (k)^2  
\,  e^{i \omega t}  \  , \label{commutf}
\end{equation}
so that
\begin{equation} 
d\nu_{A(f)} (\omega) = \omega (1 - e^{- \beta \omega})
\int \< A \, dE(k,\omega) \, A\> \, \tilde f (k)^2 \ \  
\end{equation}
for a KMS state, and similarly for a ground state.
For $f_{k,R}(x)= e^{ikx} \, \alpha_R(x)$,  $\alpha_R$ as in eq.(\ref{fr}), 
$R \to \infty$,
\begin{equation} 
d\nu_{A,f_{k,R}}(\omega) \equiv  1/R^d \ d\nu_{A(f_{k,R})}(\omega) \to   
\omega (1 - e^{- \beta \omega}) \;  
 \< A \, dE(k,\omega) \, A\>  
\label{EST}
\end{equation}
apart from a constant, as distributions in $\omega$, assuming a polynomial bound 
in $t$ for the $L^1$ norm of the commutator $\< [ A(0,0), A(y,t)] \> $
\cite{Morchio-strocchi1986}.  

For states invariant under a lattice group of space translations,
$ \<[A(x+a,0) A(y+a,t)]\>$ is periodic in $a$ and, for
$f \equiv f_{k,R}$, apart from an irrelevant constant
$$ 1/R^d \, \< [ A(\bar f), A(f,t) ] \>  \to $$
$$  \int_{\Lambda \times \Lambda} dx \, dy \, 
 \sum_n
 \< [ A(x,0), A(y+n,t) ] \>  \, e^{ik(y-x)} \, e^{ikn} = $$
$$ =  \int_{\Lambda \times \Lambda} dx \, dy \, 
 \int \< A(x,0) \, (dE(k,\omega) - dE(k, -\omega)) \,  A(y,0)  \>  
\, e^{ik(y-x)} \, e^{i \omega t}  $$
for $R \to \infty$, assuming integrability of the above 
commutator, with $\Lambda$ denoting a lattice cell 
(the integrand being periodic both in $x$ and $y$). Therefore, 
as distributions in $\omega$,  
$$
 d\nu_{A,f_{k,R}}(\omega) 
 \to \ \  {\rm const} \ \omega (1 - e^{- \beta \omega}) \times
$$
\begin{equation} 
 \int_{\Lambda \times \Lambda} dx \, dy \, 
 \< A(x,0) \, dE(k,\omega) \,  A(y,0)  \>  \, e^{ik(y-x)}  \ \ ,
\label{ESL}
\end{equation}
assuming, as above, a polynomial bound in $t$ for the commutators.

In the following, the second and fourth momenta of 
$ d\nu_{A,f_{k,R}} $ 
and their limits for $R \to \infty$ will be calculated and discussed 
in terms of powers of  $k$, for  $A = \rho(h) $,
$h(x) \geq 0$.
The one point result for the energy spectrum of plasma waves at zero momentum
will use the following

{\lem If a sequence of positive even measures $d\nu_n (\omega)$
satisfies
$$ \lim_{n} \int d \nu_n \, \omega^{2k} = c^{2k} \ \ , \ k = 0,1,2 \ \ , 
\ \ c > 0 \ \ , $$
then the sequence converges in the sense of measures to
$$ 1/2 \, (\delta (\omega - c) + \delta (\omega +c))$$
If $ \lim_{n} \int d \nu_n \, = 1 $ and 
$ \lim_{n} \int d \nu_n \, \omega^2 = 0 $, then it converges
to $\delta (\omega)$}.

\ssk\ni
Proof: $ (\omega^2 - c^2)^2  \, d \nu_n (\omega) $ are finite positive 
measures, their integrals converge to zero for $n \to \infty$ and
therefore, for any continuous bounded function $f$, 
$\int f(\omega) (\omega^2 - c^2)^2  \, d \nu_n (\omega)$ converge to zero.
Any continuous, bounded, even function $g$ can be approximated uniformly by
$$  g(c)  + f_\eps (\omega) (\omega^2 - c^2)^2 \ \ $$
with $f_\eps $ continuous and bounded; this implies 
$$ \int g(\omega) \, d\nu_n \to g(c) $$ 
and this is enough since the $d \nu_n$ are even.
The second statement follows similarly. $\Box$

\section{Equations of motion and low $k$ ex\-pan\-sions}

In the following, we analyze the energy spectrum associated to
density waves in infinite fermion systems.
As a consequence of eq.(\ref{momenti}), the calculation of the 
momenta of the corresponding measures $d \nu_{\rho(h), f} (\omega)$  
reduces to the application of the ACR relations to the time 
derivatives of $\rho$. By time translation invariance of $\Omega$,
in order to obtain the momenta up to the fourth (the fifth for $d\mu$), 
only three time derivatives must be computed.

Assumptions A), B), C) will be needed only
for the Wick polynomials of second degree, the only ones which appear 
in the time derivatives of $\rho(x) \equiv \psi^{\ast} (x) \, \psi(x) $
(eqs.(14) and (19)-(24)). 
As usual, $ : M : $ will denote the Wick ordered 
polynomial $M$ in the variables $\psi^* (x_i), \psi(x_i) $, $i = 1 \ldots N$ 
i.e the polynomial with all the $\psi^* $ on the left of 
all the $\psi$ and the sign of the corresponding permutation 
of the variables. For simplicity, we will omit the fermion field indexes.

Considerable information about the low momentum behavour of  
charge density commutators can be obtained from general principles.
The first observation is that each time derivative 
\textit{with respect to the free evolution}
of any second degree gauge invariant polynomial $P(x)$ in the fermion fields
explicitely introduces one more space derivative in the expression of
$P(x,t)$, and therefore one power of $k$ in the Fourier transform 
of any correlation function of $P$.

{\lem For gauge invariant polynomials of second degree,
$ P(x) = \sum a_{n,k}  \partial^{(n)} \psi^* (x) \partial^{(k)} \psi(x)$,
$\partial^{(n)}$ denoting a product of $n$ space derivatives,    
the commutator with the free Hamiltonian $H^0_R$, in the limit 
$R\to \infty$, is of
the form $\sum_i \partial_i Q_i (x)$, with $Q_i$ in the same class.}   
 
\ssk\ni
Proof: For the free evolution, from
\begin{equation} 
d/dt \, \psi(x) = i \lim_R \; [H^0_R , \psi(x)] = i/2m \; \Delta \psi(x)  
\label{d0psi}
\end{equation} 
and the conjugate equation it follows immediately
\begin{equation} 
d/dt \, \partial^{(n)} \psi^* (x) \, \partial^{(k)} \psi(x) = 
\label{dt0pp}
\end{equation} 
$$ = i/2m \; \partial_l 
     (- \partial_l \partial^{(n)} \psi^* (x) \partial^{(k)} \psi(x) +
        \partial^{(n)} \psi^* (x) \partial_l \partial^{(k)} \psi(x) )  
    \ \ \ \Box
$$

\ssk
A second source of powers of $k$ arises from symmetries which imply 
the vanishing of the integral of the commutators 
$ [ A(x) ,  B]  $, 
for all local variables $B$, as operator valued distributions. 
In fact, by locality of the ACR relations such commutators
have compact support, their Fourier transform is analytic 
in $k$ and vanishes at $k=0$ if the above property holds.
E.g., since the current $j_i$ generates space translations, the 
expectation value of current commutators on a space 
translation invariant state vanish at zero momentum:
\begin{equation} 
\int dx \; \Omega ( [j_i(x) , B] ) = 0
\label{commj}
\end{equation} 
From parity invariance of $\Omega$ and definite parity of $A$, 
it also follows that
$$ \< A(x) \,  d^n/dt^n \, A(y) \> $$
is even and therefore its Fourier transform, if regular enough, is  
of order $2m+2$ in $k$ if it vanishes of order $2m$.

A basic result for the following is obtained from the
commutator between the current 
$$j_i(x) = i/2m \, (\partial_i \psi^*(x) \, \psi(x)
 - \psi^*(x) \, \partial_i \psi(x)) 
$$
and the interaction hamiltonian, which gives rise to the density of force
\begin{equation} 
F_i(x) \equiv \,
:\rho(x) \,  \int dy \; \dv_{i} V_L (x-y) \, \rho(y): \ \ \ .
\label{force}
\end{equation} 
The commutator of $F_i(x)$ with any local variable is integrable, 
also in the limit $L \to \infty$ if $\dv_i V(x)$ is integrable;
it vanishes after integration in $x$ since it is odd in the 
interchange of $x$ and $y$ (a consequence of the third law of 
Newton's). The same holds for the total force, for any number 
of fields. In the case of non integrable potentials, a
subtraction is needed and a convenient relation is given by 
the following

{\lem For a space translation invariant state, if
\begin{equation} 
  \< [ :(\rho(x) - \<\rho\>) \, \dv_{i} V (x-y) \; 
           \rho(y): \, , \,  B ] \>  \ ,
\label{forcered}
\end{equation} 
is integrable in the two variables, then its integral vanishes and
\begin{equation}  
\int dx \, \lim_{L \to \infty} \int dy \,  
  \< [ :(\rho(x) - \< \rho \>) \, \dv_{i} V_L (x-y) \, \rho(y): \, , \,  B ] \>  
  = 0  \  .
\label{forceredL}
\end{equation}
In the absence of translation invariance, the same result holds
with $\<\rho\>$ replaced by $ \< \rho (x)\> $ 
and $\rho(y)$ replaced by $ (\rho(y) - \<\rho(y)\>) $ }

\ssk\ni
Proof: Using integrability, the l.h.s. of eq. (\ref{forceredL}) gives the 
integral of (\ref{forcered}), by the Lebesgue dominated convergence theorem.
On the other side, for all $L$, 
$$ \int dy \ \dv_{i} V_L (x-y) \ \<\rho\>  = 0  $$
and therefore the l.h.s. of eq.(\ref{forceredL}) is independent of the 
substitution of $ \rho(y) $ with $ \rho(y) - \< \rho \> $;
in this form, the integral vanishes for all $L$ by antisimmetry in the 
exchange of $x$ and $y$. 
The last statement follows from integrability and antisymmetry.
$\Box$

\ssk\ni
In the application below, the subtraction $ \rho(x) - \<\rho\> $ 
in Lemma 4 is a priori irrelevant for $\dv_{i} V (x)$ integrable 
(and in fact also for $\dv_{i} \dv_j V (x)$ integrable),
but crucial for the Coulomb potential, where actually 
only the subtracted terms will survive in the analysis of the 
frequency spectrum in the zero momentum limit.

\ssk\ni
The equations of motion for the density are the following 
\begin{equation}
\frac{d}{dt} \rho(x,t) = - \, \dv_{i}j_{i}(x)
\label{d1}
\end{equation}
\begin{equation}
\frac{d^{2}}{dt^{2}} \rho(x,t) =  1/m \; \dv_{i}\, (\dv_{k}S_{ki}(x)\, + 
 :\rho(x) \;  \dv_{i} \, (V_{L} \ast \rho)(x) \, :)  
\label{d2}
\end{equation}
$$
\frac{d^{3}}{dt^{3}} \rho(x,t) = + 1/m \, \dv_{i}\,\dv_{k} 
( \dot S^0_{ki}(x)\,  + \dot S^V_{ki}(x)) \; + 
$$
\begin{equation}
- 1/m \; \dv_{i}\, ( : \dv_{k} j_{k}(x) \; \dv_{i} \, (V_{L} \ast \rho)(x) \, : +  
 : \rho(x) \;  \dv_{i} \, (V_{L} \ast \dv_{k} j_{k}(x) \, : )  
\label{d3}
\end{equation}
with
\begin{equation}
S_{ki}(x)\,=\, 1/4m  \; ( \dv_{k}\psi^{\ast} \, \dv_{i}\psi  -                    
                          \dv_{k}\dv_{i}\psi^{\ast} \, \psi \,   + h.c. ) 
\ \ ,  \ \ \  \dot{S}_{ki}^{0}  =   \partial_l \, S_{lki} \ ,
\label{Ski}
\end{equation}
$ S_{lki} $ being obtained as in Lemma 3 and
\begin{equation}
 \dot{S}_{ki}^V(x)= \, -  \,( : j_k (x) \, \partial_i V_L * \rho (x) :
 + ( i \leftrightarrow k ))
\label{SkiV}
\end{equation}

\ssk
In the presence of spin, exactly the same equations hold for
$\rho, j_i$ and $S_{ij}$ replaced by the corresponding sum over
the spin index.
For N kinds of fermion fields, $\psi_l$, $l= 1 \ldots N$ with masses $m_l$, 
the same equations apply to the density operators $\rho^{(l)}$, 
the corresponding currents $j^{(l)}$ and stress tensors 
$S_{ik}^{(l)}$, with mass $m_l$ and the obvious substitution of
$\partial_i V_L * \rho$ with  
$ \sum_m \partial_i V_L^{(l,m)} * \rho^{(m)}$.
Since the potential is assumed to be regularized by the convolution with
$\gamma(x) \in {\mc S}$, and the  correlation functions of 
$\rho(x)$, $j_i(x)$, $S_{ij}(x)$ are assumed to be bounded after smearing, 
the space cutoff $L$ can be removed in all correlation functions, for all 
potentials with integrable derivatives.

Wick ordering in eqs.(\ref{d2}), (\ref{d3}), (\ref{SkiV}) can be omitted if the
partial derivatives of the potential vanish at the origin up to the third 
order. Since this leads to simplifications in the calculations (only a few 
commutators are then needed, and one can forget about Fermi fields and their 
ordering) with no substantial consequences on the results \cite{Tesi}, 
such a property will be assumed in the following for the regularization $\gamma$
of the potentials (the derivatives of $V_L$  
vanishing at the origin up to the third order if $\triangle \gamma(0) = 0 $).
Our calculations will only use the above time derivatives and the following 
basic commutators:

\begin{equation}  
[\rho(y),j_{j}(z)]\,= \,(-i)/m \; \dv_{j}^{y} (\delta (y-z) \, \rho(y) )
\label{rj}
\end{equation}
\begin{equation}
[ \rho(y),  S_{ik}(z)] \, = - i \, (
\dv_{i}^{y}(\delta(y-z)\, j_{k}(y) )\, + \,  ( i \leftrightarrow k ))
\label{rS}
\end{equation}
$$[ j_{j}(y) , S_{ik}(z)] \, = 
  \, - i/m \; ( \, \dv_{k}^{y}\delta(y-z) \,   S_{ij}(z) \, +
  ( i \leftrightarrow k ) )\, +
$$
\begin{equation}
 + i/m \; \dv_{j}^{z} (\delta(y-z)\,  S_{ki}(z) ) -
   i/4m^2 \; \dv_{i}^{y}\dv_{k}^{y} \dv_{j}^{z} (\delta(y-z)\, \rho(y))
\label{Sj}
\end{equation}
Eqs.(\ref{rj}),(\ref{rS}),(\ref{Sj}) also hold for the sum over the spin
index of the same operators. 
By eq.(\ref{momenti}), the required momenta are of the form 
$$
\int \omega^n  \, d\nu_{\rho(h),f_{k,R}}  =  
$$
\begin{equation}
= 1/R^d \, (-i)^{n+1}  
\< [ \rho(h * \bar f_{k,R} ) , \,  d^{n+1} / dt^{n+1} \, \rho(h * f_{k,R},t)] \, \> 
\, |_{t=0}  \ \ .  
\end{equation} 
For translation invariant states, 
\begin{equation}
 \int \omega^n  \, d\nu (\omega) =   
   (2 \pi)^{-d} \, \int dq  \; \tilde T_n (q) \; 
                         \tilde h (q)^2 \, \tilde \alpha_R(q-k)^2 / R^d    
\label{momT}
\end{equation} 
with
\begin{equation}
  \tilde T_n(k) \equiv   {\mc F} \, T_n (x-y) \equiv  
    (-i)^{n+1}  {\mc F}   
\, \< [ \rho(x) , \,  d^{n+1} / dt^{n+1} \, \rho(y,t)] \, \> \, |_{t=0}  \ \ .  
\label{Tn}   
\end{equation} 
From the equations of motion and boundedness of the (equal time) 
correlation functions after smearing it follows immediately that 
$\tilde T_n(k)$ is regular ($ C^\infty $ 
for potentials of fast decrease, continuous for potentials
with $\dv_i V(x)$ integrable), 
so that the limit $ R \to \infty $ of eq.(\ref{momT}) exists and 
gives $ \tilde T_n(k) $, apart from a constant times
$\tilde h(k)^2$, which is positive for small $k$.
The corresponding normalized momenta of the frequency converge therefore to
\begin{equation}
  < \omega^n (k) > = \tilde T_n(k) / \tilde T_0(k) 
\label{on}   
\end{equation} 

For states invariant under lattice translations 
(omitting for simplicity the convolution with $h$),
$$
\int \omega^n  \, d\nu_{\rho(0),f_{k,R}} (\omega) = 
$$
\begin{equation}
=  1/R^d \; (-i)^{n+1}  
\int dx \, dy \; e^{ik(y-x)} \, \alpha_R(x) \, \alpha_R(y) \,  T_n (x,y)
\label{momTL}
\end{equation} 
with $ T_n (x,y)$ given by the r.h.s. of eq.(\ref{Tn}).
For potentials with $\dv_i V(x)$ integrable, $T_n(x,y)$ 
is integrable in $y$ and its integral is periodic in $x$.
Therefore, the limit $ R \to \infty $ of eq.(\ref{momTL}) exists and 
gives, apart from an irrelevant constant, 
\begin{equation}
          1/|\Lambda| \,  \int_\Lambda dx \, \int dy \,
e^{ik(y-x)} \, T_n (x,y) \equiv \tilde T_n(k,0) \ \ , 
\label{Tn0}
\end{equation} 
with the integrand periodic in $x$, the notation referring to
the Fourier series expansion in the variable $x$.
Again, with the notation of eq.(\ref{Tn0})
\begin{equation}
  < \omega^n (k) > = \tilde T_n(k,0) / \tilde T_0(k,0) 
\label{onL}   
\end{equation} 
Eq.(\ref{rj}) gives
\begin{equation} 
  \tilde T_0(k) =  \< \rho \>  \, k^2 / m  \
\label{T0}   
\end{equation} 
\begin{equation} 
 \tilde T_0(k,0) =  1/|\Lambda| \, \int_\Lambda dx \, 
                   \< \rho(x) \>  \, k^2 / m   \equiv
                   \< \, \bar \rho \, \>  \, k^2 / m   
\label{T0L}   
\end{equation} 

Eqs.(\ref{d1}),(\ref{d2}) immediately imply that $\tilde T_2 (k)$ and
$\tilde T_2 (k,0)$ are 
infinitesimal with respect to $k^2$, as a consequence of the presence 
of space derivatives (also following from Lemma 3) and of 
the arguments following eq.(\ref{force}). This also clearly applies 
for more than one kind of fermions, for the total density 
$\rho^{tot}(x) \equiv \sum_l \rho^{(l)}(x)$.
Lemma 2 then gives a result similar to the Goldstone theorem for the 
spontaneous breaking of the Galilei group 
\cite{Swieca1967}, \cite{Morchio-strocchi1986}:

{\prop For any number of fermion fields and potentials $V^{(lm)}$
with integrable first derivatives, the energy spectral measures \\ 
$d\nu_{\rho^{tot}(h),f_{k,R}}$, 
eqs.(\ref{EST}),(\ref{ESL}),
normalized to total unit mass, 
converge as measures , for $R \to \infty$ and $k \to 0$, 
to $\delta(\omega)$}.

In the next Section the mean squared frequency is discussed
for all potentials decaying at infinity  faster than the 
Coulomb case; as a result,  Proposition 1 will be extended  
to all such potentials, more precisely to 
\textsl{all potentials with integrable second  derivatives}

\section{$ < \omega^2 (k) > $  and long range potentials}

The calculation of $\tilde T_2(k)$ and $\tilde T_2(k,0)$ 
only requires eqs.(\ref{d1}), (\ref{d2}),
(\ref{rj}), (\ref{Sj}). In fact, using time translation 
invariance of $\Omega$, 
$$
T_2 (x,y) =  - (-i)^3   \, \< [ d/dt \, \rho(x,t) , \, 
       d^2 / dt^2 \, \rho(y,t)] \> \, |_{t=0} \,  \  = 
$$
$$
   = (-i)^3/m \; \dv_i^x \dv_j^y    
     \, (   \dv_k^y  \< \, [ j_j(x) ,  S_{ki}(y)] \, \> + 
     \< \, [ j_j(x) , \,\rho(y) \, \dv_i^y \, (V_{L} \ast \rho)(y)] \, \>  ) =  
$$
$$
  =  1/m^2 \;  \dv_i^x \dv_j^y \dv_k^y  
    \, (  \dv_{k}^{x}\delta(x-y) \,  S_{ij}(y) \, + \ldots
     + 1/4 \; \dv_{i}^{x} \dv_{k}^{x} \dv_{j}^{y} \delta(x-y)\, \rho(x)) \, +
$$
$$
   +   1/m^2 \; \dv_i^x \dv_j^y    
    \int dz \, \dv_i \dv_j  V_L (y-z) \,  ( \delta(x-y)   - \delta(x-z) ) \, 
       \< \rho(y)  \rho(z) \>  \, + 
$$
\begin{equation}
   +   1/m^2 \; \dv_i^x \Delta^y 
   ( \delta(x-y) \< \rho(y) \, \dv_i V_L * \rho \, (y) \> ) 
\label{T2i}   
\end{equation} 
The dots refer to two terms obtained, as in eq.(\ref{Sj})  
by permutations of indexes. In the presence of spin, each 
$\rho$ and $S_{ij}$ operator must be summed over the spin index.
The cutoff $L$ can be removed if $ \dv_i \dv_j V (x)$ 
is integrable, as a consequence of boundedness (after smearing) 
of the two point correlation function of $\rho$. 

For translation invariant states, the last term in eq.(\ref{T2i})
vanishes by invariance under parity (or, see eq.(\ref{d2}), under 
space and time translations), so that 
\begin{equation}
  \tilde T_2(k) =   1/ m^2 \; k_i k_k \,     
    (3 \,\< S_{ik} \> \, k^2 - (\tilde G_{ik} (k) - \tilde G_{ik} (0)) ) +
  1/4 m^3 \; \< \rho \> \, k^6
\label{T2}   
\end{equation} 
with 
$$
\tilde G_{ik}(k) \equiv \, {\mc F} \, 
( \,  \dv_i \dv_k  V (x-y)  \< \rho(x) \rho(y) \> \, )   
$$
Eqs.(\ref{momT}), (\ref{T0}), (\ref{T2}) give,
for the second momentum of the frequency
\begin{equation}
   < \omega^2 (k) > \, = 
     \frac { 3 \, \< \, S_{ik} \, \> }  {m \< \, \rho \, \>}  k_i k_k 
       + \frac {1} {4 m^2} \, k^4   
      - \frac {k_i k_k} {m \< \rho \> k^2}  
          (\tilde G_{ik} (k) -  \tilde G_{ik} (0) )   
\label{F2}
\end{equation} 
In the rotation invariant case,
$$   
\< S_{ik}(x) \> =  1/d m \; \< \dv_j \psi^* \dv_j \psi \> \equiv  2/d \;
E_{kin} \, \< \rho \>
$$
Eq.(\ref{F2}) has been derived in refs. \cite{Puff1965}, \cite{Iwamoto1986}
in the framework of a finite particle system in a large box (the use of the
Coulomb potential in its derivation is not without problems, the plasma frequency
being given by a singularity at the origin in $G_{ij}(k)$).
It holds, in our framework, for all potentials with $\dv_i \dv_j V $
integrable.

For states invariant under a lattice subgroup of translations,
the last term in eq.(\ref{T2i}) vanishes by parity 
invariance after integration in $x$ on a lattice cell; 
using periodicity in $x$ and integration by parts, 
\begin{equation}
  \tilde T_2(k,0) =   1/ m^2 \; k_i k_k \, 
    (3 \,\< \, \bar S_{ik} \, \> \, k^2 - (\tilde G^0_{ik} (k) - 
    \tilde G^0_{ik} (0) )) +  1/4 m^3 \; \< \, \bar \rho \, \> \, k^6 \ \ ,
\label{T2L}   
\end{equation} 
with $\bar \rho$ and $\bar S_{ik}$ mean values over a lattice cell
and 
$$
\tilde G^0_{ik}(k) \equiv 1/|\Lambda| \int_\Lambda dx \;
    \int dy e^{ik(y-x)} \;  \dv_i \dv_k  V (x-y)  \< \rho(x) \rho(y) \>  \ \ , 
$$
so that, by eqs.(\ref{onL}),(\ref{T0L}), the mean squared frequency is 
given again by eq.(\ref{F2}), with $\rho$ and $S_{ik}$ substituted by 
their mean values and $\tilde G_{ik}$ by $\tilde G^0_{ik}$.

In all the cases, the low momentum behaviour of the mean square frequency 
depends on the decay property the potential at infinity. In fact,
$\tilde G_{ij}(k)$ and $\tilde G^0_{ij}(k)$ are $C^n$ for potentials with 
$|x|^n \dv_i \dv_j V $ integrable.
For $n = 0$,  the (immediate) extension of eqs.(\ref{F2}),(\ref{T2L}) 
to $N$ kinds of fermions proves Proposition 1 for potentials with 
integrable second derivatives. Actually, the result only depends on an 
additional power of the momentum appearing in all the commutators as a 
consequence of eq.(\ref{commj}).
For $n \geq 4$, in particular for sufficiently regular potentials 
decaying at least as $|x|^{-d -2 -\varepsilon}$, 
eq.(\ref{F2}) implies, 
using rotation invariance, 
\begin{equation}
 < \omega^2 (k) > =   
    (6/d \,  E_{kin} + C ) \, k^2 /  m \<\rho\>  \;  + O( k^4) 
\label{o2reg}
\end{equation}
The same holds for $n \geq 2$ (in particular for regular  
potentials decaying at least as $|x|^{ -d - \varepsilon}$), 
with $ O( k^4) $ replaced by $o(k^2)$.
In both cases $C$ can be written, in dimension $d=3$,
$$
  C =  1/30 \int d^3x \, \< \rho(0) \rho(x) \> 
      (x^2 \Delta  + 2 x_i x_j \dv_i \dv_j ) \, V(x)
$$
For potentials with
$ |x|^\beta \dv_i \dv_j V(x)$ integrable,  $0 < \beta <2$, the 
last term in eq.(\ref{F2}) can dominate at low momentum; 
it can be estimated as:
$$
| \tilde G_{ij} (k) - \tilde G_{ij} (0) | \leq  2 \, |k|^\beta 
 \int dx \; |x|^\beta \, |G_{ij} (x)| \ \ ,
$$
so that 
\begin{equation}
< \omega^2 (k) > \ \leq \ {\rm const} \, |k|^\beta \ \ .
\label{F2beta}
\end{equation} 

If the potential term dominates, the low momentum behaviour
of $ < \omega^2 (k) > $ is clearly independent from the addition of
any potential $W$ with $ x^2 \, \dv_i \dv_j W$ integrable;  
for potentials 
$ V(x) \sim   |x|^{-\alpha} $, $d-2 < \alpha <d$, one obtains
$$
< \omega^2 (k) > \, \sim \, |k|^\beta
$$
with $\beta = \alpha - d + 2 $.
The same results apply to states invariant under lattice
translations.

In the following Sections the fourth momentum of the frequency
is discussed, in the zero momentum limit for the Coulomb 
interaction and to the order $k^2$ for short range potentials.

\section{The plasmon spectrum in Jellium}

We will derive in this Section the one point result, 
eq.(\ref{Plasmon}), for the energy spectrum of 
density waves in the limit $k \to 0$ for the jellium model, 
in three space dimensions; 
the result will follow from the application of Lemma 4 to the 
analysis of the fourth momentum of the frequency and
of Lemma 2 to the momenta up to the fourth.

As discussed above, in the case of a Coulomb potential
an infrared cutoff is necessary in the equations of motion. 
Its removal in the first four momenta of the frequency only
requires that the two and three point correlation functions of
$\rho(x) - \<\rho(x) \>$ decay (respectively) as 
$|x_1 - x_2|^{-1-\eps}$ and
$|x_1 - x_2|^{-1-\eps} \, |x_2 - x_3|^{-1-\eps}$. 

Eqs.(\ref{F2}),(\ref{T2L})
for the second momentum of the frequency holds for potentials $V_L$
with ultraviolet and infrared regularization, eqs.(\ref{HLR}),(\ref{fr})
The first two terms in eq.(\ref{F2}), are of order $k^2$ and $k^4$;
for the last term, from integrability in $y$ of eq.
$$ 
\dv_j  V (x-y) ( \< \rho(x) \rho(y) \> - \<\rho (x)\>  \<\rho (y)\>) 
$$   
it follows that, after such a subtraction, the limit for $L \to \infty$ 
vanishes of order $k$. We are left therefore with
$$ 
   {\mc F} \; ( \, \dv_i \dv_j  V_L (x-y)  \< \rho(x) \>   \< \rho(y) \> \, )  
$$  
and therefore, in the space translation invariant case,  
\begin{equation}
  < \omega^2 (k) > = 
\lim_L  \,  k^2 \, \tilde V_L(k) \< \rho \> / m  
          + O(k)  \to
   e^2 \<\rho\>/m  
\label{F2J}
\end{equation}
for  $k \to 0$, using  $\tilde \gamma(0) = 1$.
$O(k)$ becomes $O(k^2)$ if the truncated correlation 
function of $\rho$ decays as $ |x|^{-2-\eps} $.

For lattice translation invariant states, expanding the periodic function
$\< \rho(x) \>$ in Fourier series, the corresponding term reads
$$
  \frac {k_i k_j} {m \,  \< \bar \rho \> \, k^2}
 \, \sum_n | \, \< \tilde \rho_n \> |^2 \; 
(  (k_i + n_i) (k_j + n_j) \, \tilde V_L(k+n)
  - { n_i  n_j} \, \tilde V_L(n) ) \ \ , 
$$   
$n$ ranging  over the reciprocal lattice of the translation lattice.
For $L\to \infty$, all the terms with $n \neq 0$ 
are of order $k$ and the same holds for their sum (as a consequence 
of the regularization of the potential). Therefore
$ <\omega^2 (k) > $ converges, for $k \to 0$, to 
$ e^2 \< \bar \rho\>/m $ (in fact $\bar \rho = \tilde \rho_0$),  
i.e. to the square of the plasma frequency associated to the 
mean density.

The absence of two powers of $k$ in eq.(\ref{F2J}) with respect to eq. 
(\ref{o2reg}) depends on the failure, for the Coulomb potential, of 
both the mechanisms mentioned in Sect.3, i.e., in the subtraction needed 
for the validity of Newton's third law and in the 
non-integrability in $x$ of the commutator
$$ 
 - i \, \int dy \; \<  [j_i(x) , \dv_k V(y) \rho(y)] \> =
     - 1/m \;  \dv_k  \, \dv_i \, V(x) \, \< \rho(x) \>  \ \ ,
$$
which in fact gives, when summed over equal indexes, $e^2 \< \rho \>$ 
(Gauss' law). The result for the mean squared frequency also follows 
from an analysis of the removal of the infrared cutoff in the above
commutator:
$$
 - 1/m \;  \int dx \; \dv_i \, \dv_i \, V_L(x) \, \rho(x) =  
 e^2/m \; \int dx \; (\delta(x) - \sigma_L(x))  \rho (x)  
$$
with $\sigma_L(x)$ of support near $|x| = L$ and    
$\int \sigma_L(x) \; dx =1$, so that, in all correlation functions, 
for $L \to \infty$, $ \int dx \; \sigma_L(x)  \rho (x) \to \< \bar \rho \> $ 

The fourth momentum of the frequency is
obtained, eqs.(\ref{Tn}),(\ref{on}), from the commutator between 
the r.h.s. of eqs.(\ref{d2}),(\ref{d3}). 
The complete commutator has been calculated 
in ref. \cite{Tesi}. In order to discuss the zero momentum 
limit, by eqs.(\ref{on}),(\ref{T0}),(\ref{T0L}), 
we need only terms up to the second order in $k$, i.e. the 
commutators between the last terms in equations (\ref{d2}), (\ref{d3}).
We need therefore to discuss the limit $L\to\infty$ and then
$k \to 0$ of 
\begin{equation}  
 \< \, [  \rho(x) \, \dv_i  V_{L} \ast \rho(x)  \, ,
   \, \dv_{k} j_{k}(y) \; \dv_{j} \, V_{L} \ast \rho(y)   +  
  \rho(y) \,  \dv_{j} \, V_{L} \ast \dv_{k} j_{k}(y)  ] \, \> 
\label{comm4}
\end{equation}
By applying eq.(\ref{rj}), four terms appear, each involving the three point
function of the density.
In the translation invariant case, $y$ can be fixed and $\< \rho \> $ 
subtracted in the convolutions. If we also subtract to 
the l.h.s. of the commutator a term $ \< \rho \> \; \dv_i  V_{L} \ast \rho $,
such terms are integrals, in two variables, of functions bounded by 
integrable functions  uniformly in $L$, as a consequence of the decay 
assumptions on the correlation functions of the density. 
Their limit $L \to \infty$ vanishes by Lemma 4 and therefore we can substitute 
$ \< \rho \>$ to $\rho(x)$ in the l.h.s. of the commutator.
The same argument then applies, by antisymmetry  of $\dv_i V(z) $ to the
r.h.s., for the subtraction  $\rho(y)$ $\to$ $\rho(y) - \< \rho \>$ 
We are therefore left with 
$$   \dv_i^x \, \dv_j^y
\< \, [ \, \< \rho \> \, \dv_i  V_{L} \ast \rho(x)  \, ,
     \< \rho \> \  \dv_{j} \, V_{L} \ast \dv_{k} j_{k}(y) \, ] \, \> \ \ 
$$
and eq.(\ref{rj}) gives
$$     \< \rho \>^2  \;  
       \int dw \; dz \; \triangle \,  V_{L} (x-w) 
       \;  \triangle \,  V_{L} (y-z) \,  
   (-i)/m  \; \dv_{k}^z \, \dv_{k}^{w}   
  \<  \,  \delta (w-z) \, \rho(w)  \, \> =
$$
$$
   = - i/m \;  \< \rho \>^3  \;   \int  dz \; 
     \dv_k \, \triangle \, V_{L} (x-z) \; 
     \dv_k \,  \triangle  \, V_{L} (y-z)
      \ \  ,
$$
so that, by eq.(\ref{Tn}),
$$
{\mc F} \,  T_4(k) \sim  1/m^3 \;  \< \rho \>^3  
            (k^2)^3 \; \tilde V_L^2 (k) \; \to 
     1/m^3 \;  \< \rho \>^3  \; e^4 \; k^2  \, \tilde \gamma^2(k) 
$$
for $L \to \infty$. This immediately implies
\begin{equation} 
 < \omega^4 (k) > \; = \; \tilde T_4(k) / \tilde T_0(k) \to
        \,  e^4 \, \< \rho \>^2 \,   / \, m^2
\label{F4J}
\end{equation}
for $k \to 0$, independently of the regularization $\gamma$, with
$\tilde \gamma(0) = 1$. 
By Lemma 2, eqs. (\ref{F2J}), (\ref{F4J}) imply the one point 
result, eq.(1), for the energy spectrum of density waves in the zero
momentum limit. 
The result is independent from the ultraviolet regularization of the potential
and clearly holds for all potentials with 
$k^2 \tilde V (k) \to e^2$ for $k \to 0$. 

For states invariant under lattice translations, eqs.(\ref{Tn0}),(\ref{onL}) 
apply and  we must consider the mean in $x$ and integral in $y$ of the 
commutator (\ref{comm4}). Using the cluster properties and 
applying Lemma 4, the commutator  
vanishes, for $L \to \infty$, at zero momenta, after the subtractions 
$\rho \to \rho - \< \rho(x) \> $ in the l.h.s; the remaining terms vanish, 
as above, after the same subtraction in the r.h.s., so that  
we end with the expectation value of the commutator between 
$$
    \< \rho(x)\> \, (\dv_i  V_L \ast \rho)(x)   +
             \rho(x) \, \< (\dv_i  V_L \ast \rho)(x)\>   
$$
and
$$
    \dv_{k} j_{k}(y) \; \< ( \dv_{j} \, V_{L} \ast \rho ) (y) \>   +  
  \< \rho(y) \> \,  \dv_{j} \, V_{L} \ast \dv_{k} j_{k}(y) 
$$
Using integration by parts in eq.(\ref{Tn0}) and discarding terms of 
order $ > k^2 $ (uniformly in $L$), eq.(\ref{rj}) gives
$$
\tilde T_4(k,0) \sim \lim_L  1/m^3 |\Lambda|   
\int_\Lambda dx
\int dy \, dw \, dz \,  \dv_i \dv_k V_L(x-w)   \dv_j \dv_k V_L(y-z) 
$$
\begin{equation} 
 k_i k_j  \, e^{ik(y-x)} \, 
  \rho_x \rho_w \, ( \rho_y \, (\delta(w-z) - \delta(x-z)) 
                 + \rho_z \, (\delta(x-y) - \delta(y-w))) 
\label{T4JL}
\end{equation}
with $\rho_a \equiv \< \rho(a) \>$
In the translation invariant case, only the first term appeared in 
the r.h.s., the others corresponding to an irrelevant subtraction of 
a constant in the convolutions. 
By the previous result on the second momentum of the frequency, 
the one point result for the plasma spectrum  
is equivalent to the cancellation of all the terms different 
from the mean in the Fourier expansion of the expectation value of the 
density. The expansion gives, with $\tilde \rho_n$ the Fourier coefficients
of $\rho_x$, 
$$
\lim_L \sum_{n_1 + n_2 + n_3 =0}   
\tilde \rho_{n_1} \, \tilde \rho_{n_2} \, \tilde \rho_{n_3} \
(\dv_i \dv_k \tilde V_L(-k -n_1) -   \dv_i \dv_k \tilde V_L(n_2)) 
$$
$$
(\dv_j \dv_k \tilde V_L(k -n_3) -    \dv_j \dv_k \tilde V_L(n_3))  \ \ ; 
\label{combV}
$$
by the regularity of $\tilde V(k)$ for $k \neq 0$ and the vanishing of
$\dv_j \dv_k \tilde V_L(0))$ , all the terms are of order $k$ 
uniformly in $L$, except those with $n_3 =0$; in this case, since 
$n_1 = - n_2$, the limit is non vanishing only for $n_1=n_2=n_3 = 0$. 
By the fast decrease of $\tilde V(k)$ the same holds for the sum.
The result is therefore, at the order $k^2$
\begin{equation} 
 \tilde T_4(k,0) \sim  \< \bar \rho \>^3 /m^3  \; (k^2)^3 \,
        \tilde V(-k) \, \tilde V(k)  \sim
         \, e^4 \, \< \bar \rho \>^3   \, k^2 \, / \, m^3   \ ;
\label{F4JL}
\end{equation}
as before, eq.(\ref{F4JL}) 
holds independently of the ultraviolet regularization $\gamma$,
only requiring $k^2 \tilde V (k) \; \to e^2$ for $k \to 0$, 
and implies eq.(\ref{F4J}), with $\<\rho\>$
replaced by the mean density $\< \bar \rho \>$.
In the presence of spin, the same derivation and results apply
to the charge and current operators summed over the spin index.

\section{$<\omega^4(k)>$ at the order $k^2$}

We calculate the fourth  momentum of the frequency 
at the order $k^2$, for translation and rotation invariant states, 
from eqs.(\ref{Tn}),(\ref{on}). 
From Sect.3 it follows that,
for potentials of  fast decrease, $T_4(k)$ is regular 
($C^\infty$) and at least of order $k^4$.
In ref.\cite{Tesi} $T_4(k)$  has been obtained, 
at the order $k^4$, in terms of two point functions 
$\< \rho \, S_{ik} \>$, $\< j_i \, j_k \>$ and of the three 
point function of $\rho$. 
Using identities which follow from the equation of motion, we shall 
express $T_4(k)$ at the order $k^4$, and therefore $<\omega^4(k)>$ at 
the order $k^2$, purely in terms of the three point function of the
density, for one component models with short range potentials.
By eq.(\ref{Tn}), 
\begin{equation}
   T_4 (x-y) =   (-i)   \, 
\< [ d^2 / dt^2 \rho(x,t) , \,  d^3 / dt^3 \, \rho(y,t)] \, \> 
\label{T4}
\end{equation}
at $t = 0$.
In the commutator of the r.h.s. of eqs.(\ref{d2}),(\ref{d3}) 
we drop $ \dot S^0_{ik}(x) $, 
which produces terms of order $k^6$ (also as a consequence of the 
cancellations discussed in Sect.3), use the identity
(for $A$ and $B$ of the same definite parity) 
$$ 
\< [A(x) , \dot B(y)] \> = \< [B(y) , \dot A(x)] \> =  \< [B(x) ,\dot A(y))] \> 
$$ 
and eq.(\ref{SkiV}). We obtain, using translation invariance,
$$
i m^2 \,  T_4 (x-y) = 
   - \; 2 \; \dv_{i}\, \dv_{k} \, \dv_{j} \,\dv_{m}   
     \< [ S_{ki}(x) \, , \, 
     j_j (y) \, \partial_m V * \rho (y) ] \> 
$$
\begin{equation}  
   - \; 4 \; \dv_{i} \, \dv_{j} \,\dv_{m}   
   \< [\rho(x) \,  \dv_{i}  (V \ast \rho)(x) \, , \,  
     j_j (y) \, \partial_m V * \rho (y) ] \>     
\label{T44}
\end{equation}
$$
   + \; \dv_{i}\, \dv_{j}     
     \< [\rho(x) \,  \dv_{i}  (V \ast \rho)(x) \, ,  \, 
      (\dv_{k} j_{k}(y) \, \dv_{j} (V \ast \rho)(y)  +  
       \rho(x) \,  \dv_{j}  (V \ast \dv_{k} j_{k}(y) ) \> 
$$
where $\dv$ denotes the derivation with respect to $x$.
The second and third commutators only require eq.(\ref{rj}), so that 
the result involves purely the three point function of $\rho$.
The first commutator requires eqs.(\ref{Sj}) and (\ref{rS});
dropping from eq.(\ref{Sj}) the last two terms, of higher order in $k$, 
omitting the four derivatives and integrating in $y$ we obtain, 
for the coefficient of $k_i k_j k_k k_m $,
$$
    4i   \int dz \, \dv_k \dv_m V (x-z) \, 
          1/m \, \<  S_{ij}(x) \rho (z) -  j_i (x) \,  j_j(z) \> 
$$                               
Rotation invariance, symmetry under the permutations 
$i \leftrightarrow j$ and $i \leftrightarrow k$ and summation  
over equal indices give, for the coefficient of $k^4$,
$$
   4i/5 \,  \int dz \,  V (x-z)  
          (1/m \, \<  \dv_i \dv_j S_{ij}(x) \, \rho (z)  \>   +
       \< \dv_i j_i(x) \, \dv_j j_j(z) \> ) \ \ .
$$  
The identity
$$
0 =  d/dt \< \dot \rho(x,t) \, \rho(z,t) \>|_{t=0} =   \< \dv_i j_i(x) \, \dv_j
j_j(z) \>  \; +
$$
$$
+ \; 1/m \, \< \dv_{i}\, \dv_{j} S_{ij}(x)\, + \,
\dv_i (\rho(x) \, \dv_{i}  V \ast \rho(x)) \, \rho(z) \>   
$$
then gives, for the first term in $T_4$,
\begin{equation}
  \tilde T_4^{(1)}(k) =   -4/5 \; k^4 / m^3  
    \,  \int dy \, dz \, \dv_{i}  V(y) \,  \dv_i V (z)  
               \<  \rho(0) \, \rho(y) \, \rho(z) \>   
\label{T4S1}
\end{equation}
The second and third commutators in eq.(\ref{T44})
are immediately calculated from 
eq.(\ref{rj}); in both commutators, a term with four 
space derivatives appears, of the same form as in eq.(\ref{T4S1}), 
with coefficients, respectively, $4/3$ and $1/3$.
The remaing terms only contain three or two space derivatives, and 
a Taylor expansion in $k$ of the corresponding expressions is necessary; 
the resulting contributions to $T_4(z)$ are therefore integrals
of the three point function of the density with first and second order
polynomials in $z$. Summing all the terms, we obtain, to the fourth
order in $k$, 
$$
  \tilde T_4(k) =    (-4/5 + 5/3) \; k^4 / m^3  
    \,  \int dy \, dz \, \dv_{i}  V(y) \,  \dv_i V (z)  
           \,  \<  \rho(0) \, \rho(y) \, \rho(z) \>    +
$$
\begin{equation}
 +   k_i k_j k_l k_m / m^3  \;    \int  dy \; dz \;
          Z_{ijlm} (y,z) \; \<  \rho(0) \, \rho(y) \, \rho(z) \>  \ \ ,
\label{T4S}
\end{equation}
with
$$
    Z_{ijlm} (y,z) \,  \equiv  \,
    6 \; \dv_i V(y) \, z_m \, \dv_j \dv_l V(z) +
$$
$$
  + 1/2 \; \, \dv_i \dv_k V(y) \, z_l \, z_m  \,  \dv_j \dv_k ( 2 V(z) - V(z-y))
  \ \ .
$$
By rotation invariance, the result can also be written as
\begin{equation} 
  \tilde T_4(k) =   k^4 / m^3  
    \,  \int dy \, dz \, W(y,z) \,  \<  \rho(0) \, \rho(y) \, \rho(z) \>   
\label{T4SR}
\end{equation}
with 
$$
   W(y,z) =   13/15 \ \dv_{i}  V(y) \,  \dv_i V (z) +  
 6/15 \ \dv_{i}  V(y) (z_i \triangle + 2 z \cdot \partial \, \dv_i ) V (z) +
$$
\begin{equation} 
       + 1/15 \  \dv_i \dv_k  V (y) \, \dv_j \dv_k  V (z) \,  
                    (\delta_{ij} \, y \cdot z +  y_i \, z_j + y_j \, z_i) )    
\label{W}
\end{equation}
and eq.(\ref{fourth}) follows. 

\bsk\noindent
{\bf Acknowledgements } The second author thanks F. Strocchi for
many critical discussions and suggestions.

\newpage
\addcontentsline{toc}{chapter}{Bibliografia}
\bibliographystyle{unsrt}
\bibliography{jellium}

\end{document}